\documentclass[11pt]{article}
\pdfoutput=1
\usepackage{jcappub}
\usepackage{aas_macros}
\usepackage{mathrsfs}
\usepackage{xspace}
\newlength{\wdo}
\newcommand{\stroke}[1]{{$#1$}%
\settowidth{\wdo}{${#1}$} {\kern-\wdo}%
\partialvartstrokedint}

\makeatletter
\newcommand{\fancysep}{%
  \@afterindentfalse
  {\begin{center}
    \resizebox{0.8\linewidth}{0.4ex}{{%
        \fontsize{20}{24}\usefont{U}{webo}{xl}{n}{4}}}%
  \end{center}}\@afterheading}
\makeatother

{\vskip 1.0em
\framebox[\columnwidth][r]{%
\begin{minipage}[c]{\columnwidth}%
\vspace{-1.0em}%
#1%
\end{minipage}}}
{\vskip 1.0em}

\def\XXint#1#2#3{{\setbox0=\hbox{$#1{#2#3}{\int}$}
     \vcenter{\hbox{$#2#3$}}\kern-.5\wd0}}

\newcommand{\beq}{\begin{equation}}
\newcommand{\eeq}{\end{equation}}
\newcommand{\beqa}{\begin{eqnarray}}
\newcommand{\eeqa}{\end{eqnarray}}

\newcommand{\eps}{\ensuremath{\epsilon}\xspace}

\newcommand{\nut}{\ensuremath{\nu_\tau}}
\newcommand{\bnut}{\ensuremath{\overline{\nu}_\tau}}

\newcommand{\mpl}{\ensuremath{m_\textnormal{pl}}\xspace}

\newcommand{\neff}{\ensuremath{N_\textnormal{eff}}\xspace}

\newcommand{\lnustar}{\ensuremath{L_\nu^{\star}}\xspace}

\newcommand{\phie}{\ensuremath{\phi_e}\xspace}

\newcommand{\burst}{{\sc burst}\xspace}

\newcommand{\gstar}{\ensuremath{g_\star}\xspace}

\newcommand{\num}{\ensuremath{\nu_\mu}\xspace}

\newcommand{\bnum}{\ensuremath{\overline{\nu}_\mu}\xspace}

\newcommand{\tcm}{\ensuremath{T_\textnormal{cm}}\xspace}

\newcommand{\feq}{\ensuremath{f^\textnormal{(eq)}}\xspace}
\newcommand{\feqene}{\ensuremath{f^\textnormal{(eq)}(\eps;\eta_\nu)}\xspace}

\newcommand{\nue}{{\ensuremath{\nu_{e}}}\xspace}

\newcommand{\bnue}{\ensuremath{\overline{\nu}_e}\xspace}

\newcommand{\ben}{\begin{enumerate}}
\newcommand{\een}{\end{enumerate}}

\newcommand{\deltadr}{\ensuremath{\delta_\textnormal{dr}}\xspace}
\newcommand{\bnu}{\ensuremath{\overline{\nu}}\xspace}

\newcommand{\yp}{\ensuremath{Y_{\rm P}}\xspace}
\newcommand{\dtoh}{\ensuremath{{\rm D/H}}\xspace}

\newcommand{\rhorad}{\ensuremath{\rho_{\rm rad}}\xspace}

\newcommand{\sinu}{SI$\nu$\xspace}
\newcommand{\tnu}{\ensuremath{T_\nu}\xspace}
\newcommand{\etanu}{\ensuremath{\eta_\nu}\xspace}
\newcommand{\cnue}{\ensuremath{\mathcal{C}_{\nu e}}\xspace}
\newcommand{\cepma}{\ensuremath{\mathcal{C}_{e^-e^+}}\xspace}

\title{Consequences of neutrino self-interactions for weak decoupling and big
bang nucleosynthesis}

\author[a,1]{E.\ Grohs,\note{Corresponding author.}}
\author[b]{George M.\ Fuller,}
\author[a,c]{Manibrata Sen,}

\affiliation[a]{Department of Physics, University of California, Berkeley, 
Berkeley, California 94729, USA}
\affiliation[b]{Department of Physics, University of California, San Diego, La
Jolla, California 92093-0319, USA}
\affiliation[c]{Department of Physics and Astronomy, Northwestern University, Evanston,
Illinois 60208-3112, USA}

\emailAdd{egrohs@berkeley.edu}

\abstract{We calculate and discuss the implications of neutrino
self-interactions for the physics of weak decoupling and big bang
nucleosynthesis (BBN) in the early universe. In such neutrino-sector extensions
of the standard model, neutrinos may not free-stream, yet can stay thermally
coupled to one another. Nevertheless, the neutrinos exchange energy and entropy
with the photon, electron-positron, and baryon component of the early universe
only through the ordinary weak interaction. We examine the effects of
neutrino self-interaction for the primordial helium and deuterium
abundances and $N_{\rm eff}$, a measure of relativistic energy density at
photon decoupling.  These quantities are determined in, or may be influenced
by, the physics in the weak decoupling epoch. Self-interacting neutrinos have
been invoked to address a number of anomalies, including as a possible means of
ameliorating tension in the Hubble parameter. Our calculations show that
surprisingly subtle changes in BBN accompany some of these neutrino
self-interaction schemes. 
Such minute signals require high-precision measurements, making deuterium the best abundance for BBN constraints in the models explored here.
}


\begin{document}

%\LANLppthead{LA-UR-15-20748}

\maketitle
\flushbottom

\section{Introduction}

In this paper we report detailed calculations of the evolution of the early
universe when self-interacting neutrino standard model extensions are adopted.
Though many properties of the neutrinos have been measured, including the
mass-squared differences and three of the four parameters in the unitary
transformation between the neutrino weak interaction (flavor) basis and the
mass basis \cite{PDG}, much remains unexplored and mysterious in this sector of
particle physics. Indeed, many extensions of the standard model in the neutrino
sector have been proposed, and for a variety of purposes. Standard model
extensions that posit interactions among the neutrinos only -- sometimes termed
self-interactions or \lq\lq secret\rq\rq\ interactions -- are notoriously
difficult to constrain in experiments. However, two astrophysical venues
perhaps offer a window into this physics. Both the early universe
\cite{1987PhLB..192...65R} and collapsing or merging compact objects
\cite{1988ApJ...332..826F} at some point have their energy and dynamics
influenced by neutrino-neutrino scattering. The former venue is a promising
test bed for this physics because of the advent of high precision cosmology;
the latter because of the burgeoning capabilities in multi-messenger astronomy.
Cosmological invocations of Self-Interacting Neutrino (\sinu) models include
attempts to: reconcile sterile neutrino interpretations of neutrino anomalies
with cosmological bounds
\cite{2014PhRvL.112c1803D,2014arXiv1411.1071C,2013PhRvD..87g3006S,2017JCAP...07..038F};
produce sterile-neutrino dark matter \cite{2019PhRvD.100b3533J,
2019arXiv191004901D}; suppress active neutrino free-streaming
\cite{1987PhLB..192...65R}; and act as a principal participant in schemes
crafted to reconcile discordant measurements of the Hubble parameter
\cite{Cyr-Racine:2013jua,2019arXiv190200534K}. \sinu can also put constraints
on models of inflation \cite{Barenboim:2019tux,2019arXiv191213488B} and
influence flavor evolution of dense neutrino streams in compact objects
\cite{Blennow:2008er,Das:2017iuj,Dighe:2017sur}.

There are many standard model extensions in the neutrino sector which
incorporate neutrino self-interactions. Some of these, for example, singlet
Majoron and Familon models
\cite{1981PhLB...99..411G,1981NuPhB.193..297G,1985PhLB..159...57G} envision a
spontaneously broken symmetry with a Goldstone boson which mediates neutrino
self-interactions. In these models, exchange of this boson engenders
flavor-changing interactions among the neutrinos with interaction strengths
that can be many orders of magnitude stronger than the standard model weak
interaction. Here we consider a scalar-exchange model based off of Eq.\ (2) in Ref.\ \cite{2019arXiv191004901D}.  The
interaction Lagrangian $\mathcal{L}_{\rm int}$ has effective
couplings $G_{i j}$ between neutrino species $i$ and $j$ 
\begin{equation}
  \mathcal{L}_{\rm int} = g_{ij}\overline{\nu^c_{iL}}\,\nu_{jL}\varphi +
  g_{ij}\overline{\nu_{iL}}\,\nu^c_{jL}\varphi,
  \quad
  G_{i j} = {{g_{i j}^2 }\over{ m_{\varphi}^2}},\label{eq:l_int}
\end{equation}
where $\nu_L$ is a left-handed-neutrino Dirac spinor and
$m_{\varphi}$ is the rest mass of the scalar exchange particle $\varphi$.
The superscript $c$ on the Dirac spinor indicates charge conjugation.
With the model in Eq.\ \eqref{eq:l_int}, neutrinos cannot undergo spin-flips at order $g^2$.  Spin-flips are possible at order $g^3$, but such a mechanism requires other features of an ultraviolet-complete theory, e.g., Eq.\ (1.3) in Ref.\ \cite{Berryman:2018ogk}.
The overall coupling $G_{i j}$ is analogous to the Fermi constant $G_{\rm F}$
in the standard weak interaction, so that typical cross sections in our adopted
schematic self-interaction model will scale with energy as $\sigma_{i j} \sim
G_{i j}^2\,E_\nu^2$. 
References \cite{Berryman:2018ogk,Kelly:2019wow} study how to embed the operator in Eq.\ \eqref{eq:l_int} into UV-complete models.

In this work we will consider only processes which couple neutrinos to other
neutrinos, albeit with flavor changing currents.  Specifically, we include $\nu\nu$ and $\bnu\bnu$ $s$-channel processes and $\nu\overline{\nu}$ $t$-channel processes, where the later manifests in lepton-asymmetric systems.  We will not consider
processes which convert a neutrino into an anti-neutrino. It is most convenient to represent the
couplings in our generic model in flavor-basis matrix form: 
\begin{equation}\label{eq:g_matrix}
\left[ g_{ i j}\right]\, =\,
\begin{pmatrix} 
      g_{e e} & g_{e \mu} & g_{e \tau} \\
      g_{\mu e} & g_{\mu \mu} & g_{\mu \tau} \\
      g_{\tau e} & g_{\tau \mu} & g_{\tau \tau} \\
   \end{pmatrix}\ \Rightarrow\ 
   g \begin{pmatrix} 
      1 & 1 & 1 \\
      1 & 1 & 1 \\
      1 & 1 & 1 \\
   \end{pmatrix}.
   \end{equation}
Here the flavor off-diagonal entries produce flavor-changing couplings. The
second matrix represents the coupling scheme we adopt for our early universe
weak decoupling and Big Bang Nucleosynthesis (BBN) calculations. We employ an
overall coupling strength $g$ and equal strength in all neutrino-neutrino
scattering channels, including those which change flavor. Our generic model
then has two independently specifiable quantities: $g$ and $m_{\varphi}$. 

We will further restrict the range of neutrino self-interaction parameters we
actually simulate by restricting the mass of the scalar $m_\varphi$ to be much
larger than the range of temperatures $T$ that our weak decoupling neutrino
transport and nuclear reaction calculations cover. In other words, we will
assume $m_\varphi \gg T$. The temperature range of our calculations span the
entire weak decoupling and BBN epoch. Generously, and therefore conservatively,
we take this range to be $30\,{\rm MeV} > T > 1\,{\rm keV}$. Restricting the
scalar mass to be $m_{\varphi} \gg 30\,{\rm MeV}$ means that we will not have
to be concerned with on-shell production and population of these particles
during weak decoupling. Moreover, though the scalar particles may be produced
in significant abundance early on, for example at temperatures $T >
m_{\varphi}$, we will consider only coupling strengths $g$ sufficiently large
to keep these particles in equilibrium with the neutrino component. This
guarantees that the scalar particles will have disappeared by the beginning
time of our simulations. Such relatively large couplings also preclude
out-of-equilibrium scalar particle decays and concomitant entropy generation
during weak decoupling and BBN.  
We note, however, that cosmology does not provide the only constraint on \sinu models \cite{2019PhRvL.123s1102B}.

Finally, we assume that the overall coupling $g$ is large enough to produce
efficient energy and flavor exchange among the neutrinos.  We compare the
self-interacting neutrino scattering rate, $\Gamma_{\rm SI}$, to the Hubble
expansion rate, $H$, to estimate when the neutrinos decouple from themselves
\beq
  \Gamma_{\rm SI}\sim H\implies \left(\frac{g^2}{m_\varphi^2}\right)^2\langle
  E^2\rangle n_\nu\sim\sqrt{\frac{8\pi}{3}\frac{\pi^2}{30}\gstar}\frac{T^2}{\mpl},
\eeq
where $\langle E^2\rangle$ is the average square energy, $n_\nu$ is the number
density of neutrinos, \gstar is the relativistic degrees of freedom
\cite{1990eaun.book.....K}, and \mpl is the Planck mass.  If we approximate
$\langle E^2\rangle\approx T^2$ and use Fermi-Dirac (FD) statistics for 3
flavors of neutrinos to deduce $n_\nu = 3\times3\zeta(3)T^3/(4\pi^2)$, we can
solve for $g$
\beq
  g \sim 5.2\times10^{-2}\left(\frac{m_\varphi}{30\,{\rm MeV}}\right)
  \left(\frac{1\,{\rm keV}}{T}\right)^{3/4}\left(\frac{\gstar}{3.36}\right)^{1/8}.
\eeq
For a decoupling temperature $T\simeq1\,{\rm keV}$, we estimate an effective
coupling $G_{\rm eff}\equiv g^2/m_\varphi^2\simeq3.0\times10^{-6}\,{\rm
MeV}^{-2}$, in line with the models proposed in Ref.\
\cite{2019arXiv190200534K} to alleviate the Hubble parameter tension
\cite{2019ApJ...876...85R}.
For $m_\varphi=30\,{\rm MeV}$, kaon decay \cite{2016PhRvD..93e3007P,2018PhRvD..97g5030B} rules out a universal coupling in this range \cite{2019PhRvL.123s1102B}, implying the mediator mass must be larger than $\sim100\,{\rm MeV}$.
With our assumption that $m_\varphi \gg T$, this
\lq\lq self-equilibrium\rq\rq\ in the neutrino component is effected via
two-to-two neutrino scattering processes on timescales short compared to
standard charged- and neutral-current weak interaction times, and for the
entire temperature range employed in our calculations.

The neutrino component will be essentially instantaneously self-coupled and
energy- and flavor-equilibrated in this model. However, the neutrinos will
exchange energy and entropy with the photon-electron-positron-baryon plasma
only through the ordinary standard model weak interactions. This sets up a
formidable transport problem wherein we must follow self-consistently the flow
of energy and entropy between the various components while simultaneously
calculating the strong, electromagnetic and weak interactions that determine
the quantities that can be probed, i.e., the light element abundances and relic
neutrino energy density. The latter quantity derives from the inferred
radiation energy density $\rho_{\rm rad}$ at the photon decoupling epoch, $T
\approx 0.2\,{\rm eV}$, and is parameterized by $N_{\rm eff}$, defined through   
\beq\label{eq:neff}
  \rhorad = \left[2+\frac{7}{4}
  \left(\frac{4}{11}\right)^{4/3}\neff\right]\frac{\pi^2}{30}T^4.
\eeq

The outline of this paper is as follows.  To describe the \sinu scenario, it
will prove helpful to give a brief review of BBN in Sec.\ \ref{sec:BBN}.
Subsections \ref{ssec:BBN_physics} and \ref{ssec:BBN_numerics} describes the
physics and numerics of BBN with particular emphasis on the evolution of the
neutrino seas.  In Sec.\ \ref{sec:no_deg}, we present our first \sinu results
with neutrino-anti-neutrino symmetric initial conditions.  Subsection
\ref{ssec:no_deg_physics} specifies the physics and computational model of
\sinu, and Subsection \ref{ssec:no_deg_results} gives results.  To further
investigate self-interactions during BBN, we provide two extensions to standard
BBN with an addition of dark radiation and neutrino-anti-neutrino asymmetric
initial conditions in Secs.\ \ref{sec:dr} and \ref{sec:w_deg}.  We summarize
our results and discuss further implications of self-interacting neutrinos in
Sec.\ \ref{sec:concl}.  Throughout this paper we set $c=\hbar=k_B=1$.

\section{Overview of Big Bang Nucleosynthesis}
\label{sec:BBN}

We direct the reader to Refs.\
\cite{WFH:1967,1990eaun.book.....K,2002PhR...370..333D,2008cosm.book.....W}
for a treatment on the basic physics
of BBN, and Ref.\ \cite{Cyburt:2016RMP} for an overview of the current status
of the field.  We give a brief review on the relevant topics of standard BBN here to
contrast the standard scenario with that of \sinu.

\subsection{Physics}
\label{ssec:BBN_physics}

Big Bang Nucleosynthesis is a succession of out-of-equilibrium freeze-outs.
The first freeze-out epoch is weak decoupling, when the neutrinos decouple from
the charged leptons and themselves.  Weak decoupling commences at temperature
scales of a few MeV.  The weak interaction mediates an exchange of energy from
the electrons and positrons into the neutrino seas
\begin{align}
  \nu + e^\pm \leftrightarrow \nu + e^\pm,\label{eq:wd1}\\
  \nu + \bnu \leftrightarrow e^- + e^+.
\end{align}
Both of the above processes include charged-current and neutral-current
interactions.  Electron-flavor neutrinos will experience both charged and
neutral-current interactions, while $\mu$ and $\tau$-flavor only experience
neutral-current.  As a result, more heat flows into the electron-flavor sector
than either the $\mu$ or $\tau$.  Three other neutral-current interactions
redistribute energy among the neutrino seas
\begin{align}
  \nu_i + \nu_j \leftrightarrow \nu_i + \nu_j,\\
  \nu_i + \bnu_j \leftrightarrow \nu_i + \bnu_j,\\
  \nu_i + \bnu_i \leftrightarrow \nu_j + \bnu_j.\label{eq:wd5}
\end{align}
The above three interactions lead to a ``flavor-equilibration'' (see Figs.\ [7]
and [8] in Ref.\ \cite{Trans_BBN}) among the electron and $\mu/\tau$ flavors.

Weak decoupling concludes at a temperature of a few 10's of keV.  The neutrinos
will also decouple from the baryons during the Weak Freeze-Out (WFO) epoch.
WFO and weak decoupling have a significant overlap in time.  However, due to
the large number of neutrinos compared to baryons, WFO and weak decoupling are
not simultaneous.  We monitor the change in the neutron-to-proton ratio of the
baryons by calculating six neutron-to-proton rates, schematically given as
\begin{align}
  \nue + n &\leftrightarrow p + e^-,\\
  e^+ + n &\leftrightarrow p +\bnue,\\
  n &\leftrightarrow p + \bnue + e^-.
\end{align}
We do not consider the energy and entropy flow from the baryon component into
the neutrino seas as this is a small change compared to the entropies of the
neutrino seas and plasma.  

As the temperature continues to decrease, the electron-positron pairs will
annihilate into photons during the eponymous electron-positron
annihilation epoch.  The last freeze-out epoch is Nuclear Freeze-Out (NFO),
when the nuclear abundances go out of nuclear statistical equilibrium and
obtain their primordial values.  For our purposes, the physics in
$e^\pm$-annihilation and NFO are identical in the standard and \sinu scenarios.
Although the physics is distinct, the preceding four freeze-out epochs are not
temporally distinct.  They overlap with one another and require computational
tools to model the physics.

\subsection{Numerics}
\label{ssec:BBN_numerics}

At high temperatures, weak interactions keep the neutrinos in thermal and
chemical equilibrium with the electromagnetic plasma, i.e., $\tnu=T$.  The time
derivative of the plasma temperature $T$ is based off of Eq.\ (D28) in \cite{letsgoeu2}
\beq\label{eq:dtempdt}
  \frac{dT}{dt} = -3H\left(\frac{d\rho_{\rm pl}}{dT}\right)^{-1}\left(\rho_{\rm pl} + P_{\rm pl}
    - \frac{1}{3H}\frac{dQ}{dt}\biggr|_{T}\right)
\eeq
where $\rho_{\rm pl}$ is the energy density of the plasma (less baryons),
$P_{\rm pl}$ is the pressure exerted by all plasma components, $dQ/dt|_T$ is
the plasma heat gained/lost from microphysics, and $d\rho_{\rm pl}/dT$ is the
temperature derivative of the plasma components (including baryons)
\cite{changing_weak}.  When calculating the electromagnetic energy densities,
we implement finite temperature QED effects to the electromagnetic equation of
state as explained in \cite{xmelec,1994PhRvD..49..611H,1997PhRvD..56.5123F}
(see Ref.\ \cite{2019arXiv191104504B} for a detailed analysis of the
finite-temperature QED problem in the early universe).  At high temperatures,
the plasma consists of photons, electrons, positrons, neutrinos, and baryons.

We begin the weak decoupling process at 10 MeV by dissociating the
electromagnetic components from the neutrino seas.  The electromagnetic plasma
temperature continues to evolve according to Eq.\ \eqref{eq:dtempdt} without
neutrinos.  In the standard scenario, we evolve the neutrino occupation
fractions $f_i(\eps),\,i=e,\mu,\tau$ using a weak-interaction collision
operator
\begin{align}
  \mathcal{C}_j[f_i(\eps_1)]\biggr|_{\eps_1=E_1/\tcm}
  = \frac{1}{2E_1}\int&\frac{d^3p_2}{(2\pi)^3 2E_2}
  \frac{d^3p_3}{(2\pi)^3 2E_3}\frac{d^3p_4}{(2\pi)^3 2E_4}
  \langle|\mathcal{M}_j|^2\rangle\delta^{(4)}(P_1 + P_2 - P_3 - P_4)\nonumber\\
  &\times\left[(1-f^{(1)})(1-f^{(2)})f^{(3)}f^{(4)}
  - f^{(1)}f^{(2)}(1-f^{(3)})(1-f^{(4)})\right],\label{eq:coll}
\end{align}
where $j$ indicates the collision type corresponding to a reaction in Eqs.\
\eqref{eq:wd1} -- \eqref{eq:wd5}. $P_k$ is the four momentum for particle $k$
and $\langle|\mathcal{M}_j|^2\rangle$ is the amplitude for reaction $j$ (see
Table I in Ref.\ \cite{Trans_BBN}).  We use the shorthand notation
$f^{(k)}\equiv f(\eps_k),\,\,k=1,2,3,4$, to denote the fermion occupation
numbers for either neutrinos or charged leptons, where the latter is always a FD
distribution
\beq
  f_{e^\pm}(\eps)
  = \frac{1}{\displaystyle \exp\left(\eps\frac{\tcm}{T} \mp \phie\right) + 1},
\eeq
and \phie is the electron degeneracy parameter defined below.  There exists an
implicit subscript on $f^{(k)}$ in Eq.\ \eqref{eq:coll} to denote either the
flavor (neutrino or antineutrino) or the charged lepton (electron or positron)
depending on the collision $j$.  We can calculate the heat flow from the plasma
to the neutrino seas by integrating over the collision terms involving the
charged leptons
\beq
  \frac{\partial\rho_\nu}{\partial t}\biggr|_a = \frac{\tcm^4}{2\pi^2}\sum\limits_{i=1}^6
  \int d\eps\,\eps^3\{\cnue[f_i(\eps)] + \cepma[f_i(\eps)]\}.\label{eq:nu_heat_flow}
\eeq
Equation \eqref{eq:nu_heat_flow} communicates the change to the plasma using the
$dQ/dt|_T$ term in Eq.\ \eqref{eq:dtempdt}
\footnote{We have switched which variable we hold constant in our derivatives
with respect to time, either the plasma temperature $T$ or the scale factor
$a$.  Although the distinction is important in a thermodynamic sense, $T$ can
be parameterized using $a$ implying $d\rho_\nu/dt|_a = d\rho_\nu/dt|_T$.}.
Therefore, the final expression for the heat gain/lost due to microphysics is
\beq
  \frac{dQ}{dt}\biggr|_T = \frac{\partial\rho_{\rm pl}}{\partial t}\biggr|_{\rm Nuc} -
  \frac{\partial\rho_\nu}{\partial t}\biggr|_a,
\eeq
where the first term is the heat gained from nucleosynthesis, and the minus
sign indicates the plasma loses energy to the neutrino seas during weak
decoupling.

The only terms in the evolution for the neutrino occupation numbers are the
collision operators, i.e.,
\beq
  \frac{df_i(\eps)}{dt} = \sum\limits_j\mathcal{C}_j[f_i(\eps)].
\eeq
We only evolve the occupation numbers of the neutrinos after the universe has
reached a temperature of 10 MeV.  When the neutrinos decouple in the standard
scenario, they do not follow an equilibrium distribution and so there is no
need to evolve temperatures or degeneracy parameters of the neutrino
components.  \tcm acts as an energy scale for the neutrinos so that we may
calculate energy density quantities, e.g., Eq.\ \eqref{eq:nu_heat_flow}.

Equation \eqref{eq:dtempdt} is derived from conservation of energy
\cite{letsgoeu2}, but does not conserve particle number.  However, we do
implement conservation of charge by evolving the electron degeneracy parameter
\beq
  \phie\equiv\frac{\mu_{e^-}}{T},
\eeq
for electron chemical potential $\mu_{e^-}$.  The positron chemical potential
is equal in magnitude and opposite in sign if we assume chemical equilibrium
between the electrons and positrons.  We evolve \phie with time by ensuring
electrons and positrons maintain FD distributions and the difference of the two
number densities is equal to the number density of protons (both free and those
bound in nuclei)
\beq\label{eq:chargecons}
  n_{e^-} - n_{e^+} = n_b\sum\limits_i Z_iY_i,
\eeq
for baryon number density $n_b$, atomic numbers $Z_i$, and nuclear abundances
$Y_i\equiv n_i/n_b$.  We do not explicitly evolve the degeneracy parameters for
the nuclei, but instead evolve their abundances $Y_i$ through the nuclear
reaction network with an example equation
\beq\label{eq:dydt}
  \frac{dY_i}{dt} = \sum\limits_{j,k,l}\left\{\frac{Y_k^{N_k}Y_l^{N_l}}{N_k!N_l!}[kl]_{ij}
  -\frac{Y_i^{N_i}Y_j^{N_j}}{N_i!N_j!}[ij]_{kl}\right\},
\eeq
where $N_{i,j,k,l}$ are the number of each nucleus in the reaction $i(j,k)l$,
$[kl]_{ij}$ is the reaction rate coefficient to create nucleus $i$, and
$[ij]_{kl}$ is the reverse reaction rate coefficient \cite{letsgoeu2}.  The
comoving number of protons changes as the weak-charged-current reactions change
neutrons into protons, and vice-versa.  We use Eqs.\ [19-24] in Ref.\
\cite{WFO_approx} to calculate the rates, which are input into Eq.\
\eqref{eq:dydt}.  Our rate expressions are single integrals over the neutrino
energy.  We can use either FD expressions or out-of-equilibrium occupation
fractions to characterize the neutrino distribution.

The last quantity to evolve is the scale factor with the Friedmann Equation
\beq\label{eq:hub}
  H\equiv\frac{1}{a}\frac{da}{dt} = \sqrt{\frac{8\pi}{3\mpl^2}\rho_{\rm tot}},
\eeq
which defines the Hubble expansion rate $H$.  \mpl is the Planck mass, and
$\rho_{\rm tot}$ is the total energy density of the universe: photons,
electrons, positrons, neutrinos, baryons/nuclei, dark radiation, dark matter,
and dark energy.  In practice, we ignore the contribution from dark matter and
dark energy as they are negligible during the BBN epoch.

\subsubsection{Computational Code}

Public BBN codes exist, see Refs.\
\cite{2008CoPhC.178..956P,2012CoPhC.183.1822A,2018PhR...754....1P}.  We use our
code \burst based off of Refs.\ \cite{GFKP-5pts:2014mn,Trans_BBN}.  We
integrate the coupled ordinary differential equations for $T$, \phie, $a$, $\{Y_i\}$, $\{f_i(\eps_j)\}$
with respect to the coordinate time $t$.  Our calculations commence at $\tcm =
30\,{\rm MeV}$ and terminate at $\tcm=1\,{\rm keV}$.  We begin calculating the
neutrino collision integrals at $\tcm=10\,{\rm MeV}$.  We integrate the nuclear
reaction network at all times.

\subsubsection{Results in the Standard Scenario}

In the standard scenario, we find $\neff = 3.044$, and the following abundances
\begin{align}
  & \yp = 0.2478,\\
  & \dtoh = 2.624\times10^{-5},
\end{align}
for a baryon density $\omega_b=0.02226$, consistent with Ref.\
\cite{2018arXiv180706209P}, and a neutron lifetime $\tau_n=885.1\,{\rm s}$.
Our value for primordial lithium is $^7{\rm Li}/{\rm H} = 4.375\times10^{-10}$.
We caution that the abundances are dependent on the nuclear reaction rates, and
so the above quantities should be taken as a baseline when compared to the
\sinu scenario, and not absolute predictions.

\section{Lepton Symmetric Initial Conditions}
\label{sec:no_deg}

\subsection{Neutrino transport with Self-Interactions}
\label{ssec:no_deg_physics}

With self-interactions, we need to differentiate between two different
epochs/phenomena.  We will term the process of neutrinos decoupling from the
electromagnetic plasma ``weak decoupling,'' in accordance with the terminology
used in the standard scenario.  When neutrinos decouple from each other, we
will call this process ``neutrino decoupling.''  Weak and neutrino decoupling
are contemporaneous in the standard model, but occur on vastly different
time/energy scales in the \sinu scenario.  In our model, we assume that the
self-interactions keep the neutrinos in a FD distribution throughout weak
decoupling.  This neutrino distribution differs from the electron/positron FD
distribution in both number and energy density.  For lepton-symmetric initial
conditions, the FD distribution is identical for neutrinos and anti-neutrinos
and we characterize it using a temperature \tnu and degeneracy quantity
\beq
  \etanu = \frac{\mu_\nu}{\tnu}
\eeq
for neutrino chemical potential $\mu_\nu$.  As the universe evolves through
weak decoupling, there will be a transfer of both energy and particle number to
the neutrino seas through the weak interactions.  The self interactions only
thermalize the neutrino seas by exchanging energy among the constituent
particles. These interactions neither create nor destroy neutrino number
through communication with the electromagnetic plasma. As a result, energy and
number are conserved in the \sinu scenario, unlike how we conserve energy and
charge for the electromagnetic plasma components in Eqs.\ \eqref{eq:dtempdt}
and \eqref{eq:chargecons}.  We must evolve \tnu and \etanu using the
definitions of energy and number density
\beq\label{eq:nrho}
  \left.
  \begin{array}{l}
  \displaystyle n(\tnu,\etanu) = N_\nu\frac{\tnu^3}{2\pi^2}\int d\eps\, \eps^2\feqene \\
  \displaystyle \rho(\tnu,\etanu) = N_\nu\frac{\tnu^4}{2\pi^2}\int d\eps\, \eps^3\feqene
  \end{array}\right\}
  \implies
  \begin{cases}
    \tnu = \tnu(n,\rho)\\
    \etanu = \etanu(n,\rho)
  \end{cases},
\eeq
where we write \tnu and \etanu as functions of $n$ and $\rho$, and the FD
expression \feqene is
\beq
  \feqene = \frac{1}{e^{\eps-\etanu}+1}.
\eeq
In Eq.\ \eqref{eq:nrho}, $N_\nu$ is the number of particle species and is equal
to 6 (\nue, \bnue, \num, \bnum, \nut, \bnut) in the non-degenerate \sinu
scenario.  If we add energy or number to the neutrino seas, we change both \tnu
and \etanu via the definitions of energy and number density
\beq
  \frac{d\chi}{dt} = \frac{\partial\chi}{\partial \tnu}\frac{d\tnu}{dt}
  + \frac{\partial\chi}{\partial\etanu}\frac{d\etanu}{dt},
\eeq
where $\chi$ denotes either $\rho$ or $n$.
Therefore, we find the time
derivative of \tnu to be
\beq\label{eq:dtnudt}
  \frac{d\tnu}{dt} = -H\tnu + \tnu\frac{\displaystyle n,_{\eta}\frac{\partial\rho}{\partial t}\biggr|_a
  - 3\tnu n\frac{\partial n}{\partial t}\biggr|_a}
  {\displaystyle 4\rho n,_\eta - 9\tnu n^2},
\eeq
and the time derivative of \etanu
\beq\label{eq:detanudt}
  \frac{d\etanu}{dt} = \frac{\displaystyle 4\rho\frac{\partial n}{\partial t}\biggr|_a
  - 3n\frac{\partial\rho}{\partial t}\biggr|_a}
  {\displaystyle 4\rho n,_\eta - 9\tnu n^2},
\eeq
where $\partial n/\partial t|_a$ is the number density added from
out-of-equilibrium weak decoupling, $\partial\rho/\partial t|_a$ is the energy
density added, and $n,_\eta$ is the following expression
\beq\label{eq:n_eta}
  n,_\eta = N_\nu\frac{\tnu^3}{\pi^2}\int d\eps\,\eps\feqene.
\eeq

We have written \tnu and \etanu as functions of energy and number density in
Eq.\ \eqref{eq:nrho}.  Although $\rho$ and $n$ have familiar meanings in an
out-of-equilibrium system, temperature and chemical potential (i.e., degeneracy
parameter) do not have the same thermodynamic meaning as they would in an
equilibrium system.  Simply stated, \tnu and \etanu are parameters for energy
and number distributions.  Equations \eqref{eq:dtnudt} and \eqref{eq:detanudt}
give the equations of motion for the two parameters and are consistent with the
definitions of $n$ and $\rho$ in Eq.\ \eqref{eq:nrho}.  The electromagnetic
plasma acts as a heat source for the neutrinos, but the lack of a tight-coupling
between the two systems precludes the equivalence of \tnu and \etanu with the
thermodynamic variables of temperature and degeneracy parameter of the grand
canonical ensemble.  Nevertheless, we will colloquially refer to \tnu and
\etanu as temperature and degeneracy parameter, respectively, throughout this
manuscript.

There are two weak interaction collision terms we use in the \sinu scenario
\begin{align}
  \nu + e^\pm \leftrightarrow \nu + e^\pm,\label{eq:nue}\\
  \nu + \bnu \leftrightarrow e^- + e^+,\label{eq:nuepma}
\end{align}
with collision terms \cnue and \cepma, respectively.  \cnue and \cepma are the
same collision operators for the standard scenario of out-of-equilibrium
neutrino spectra (see Appendix C in Ref.\ \cite{Trans_BBN}).  For our purposes,
we do not evolve the individual occupation fractions, $f_{\nu_i}(\eps)$, of the
neutrino spectra.  Instead, we use the FD expression \feqene for both terms and
integrate over \eps to find the total added neutrino number and energy density
for a given time step.  Note that \cnue preserves the neutrino number and
changes the energy, while \cepma changes both number and energy.  Therefore,
the number and energy density added during the out-of-equilibrium weak
decoupling is
\begin{align}
  \frac{\partial n}{\partial t}\biggr|_a &= \frac{\tnu^3}{2\pi^2}\sum\limits_{i=1}^6
  \int d\eps\,\eps^2\cepma[\feqene],\label{eq:sinu_num_flow}\\
  \frac{\partial\rho}{\partial t}\biggr|_a &= \frac{\tnu^4}{2\pi^2}\sum\limits_{i=1}^6
  \int d\eps\,\eps^3\{\cnue[\feqene] + \cepma[\feqene]\}.\label{eq:sinu_heat_flow}
\end{align}
Equation \eqref{eq:sinu_heat_flow} is a copy of Eq.\ \eqref{eq:nu_heat_flow}
except we use \tnu instead of \tcm for the energy scale.

There is neither number nor energy density added to the neutrino seas after
weak decoupling concludes.  At this point, Eqs.\ \eqref{eq:dtnudt} and
\eqref{eq:detanudt} reduce to
\begin{align}
  &\frac{d\tnu}{dt} = -H\tnu,\\
  &\frac{d\etanu}{dt} = 0,
\end{align}
which are the free-streaming conditions for relativistic particles.

\subsection{Results}
\label{ssec:no_deg_results}

\subsubsection{Neutrino Spectra}

\begin{figure}[ht]
   \begin{center}
   \includegraphics[scale=0.6]{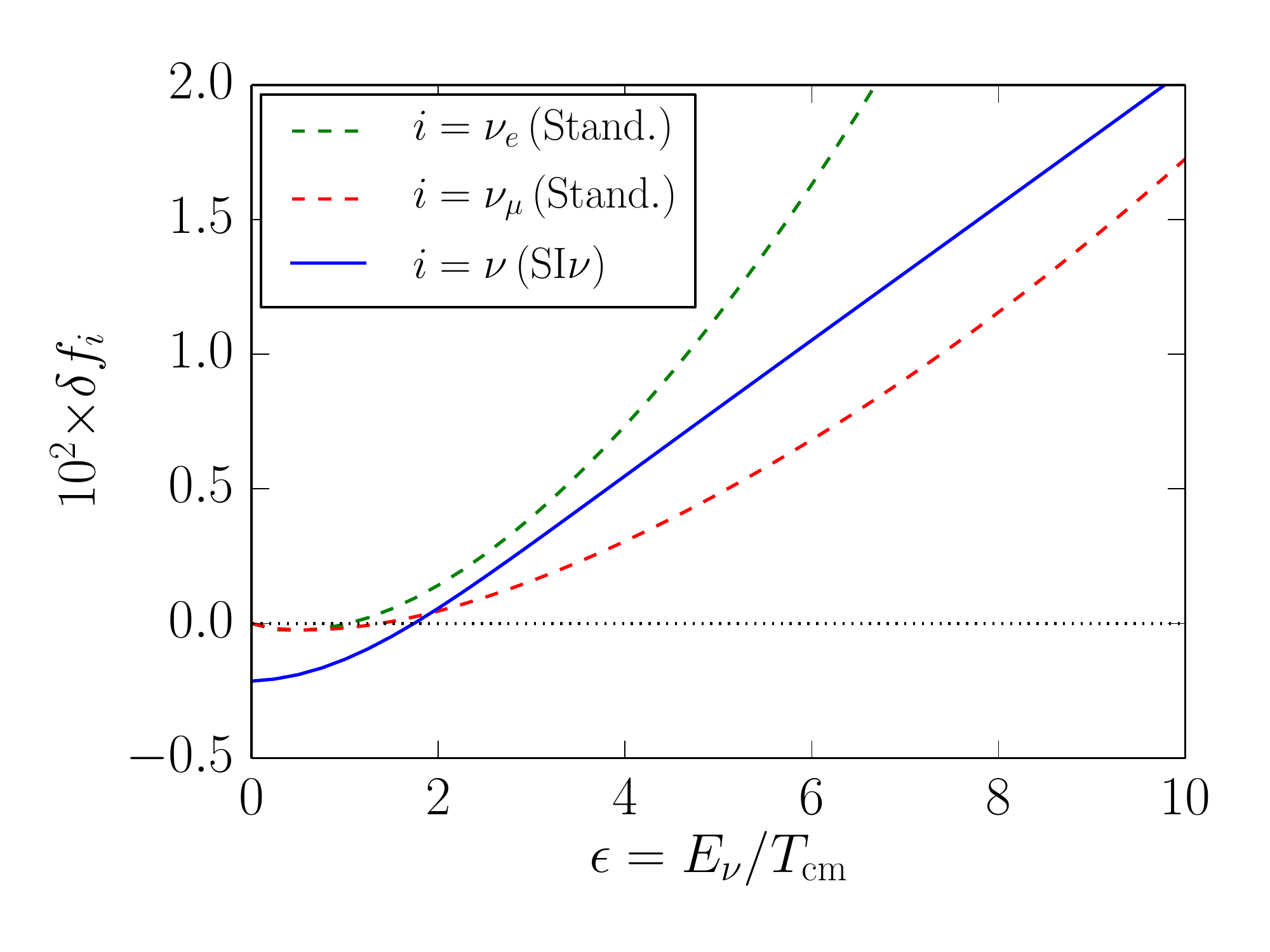}
   \caption{\label{fig:docc} Freeze-out neutrino spectra in the standard
   scenario (dashed) and the \sinu scenario (solid) plotted against \eps.
   In the standard scenario of Boltzmann transport, the green dashed curve is the
   $e$ flavor and the red dashed curve is either $\mu$ or $\tau$.  Vertical axis
   is the relative change to a non-degenerate FD spectrum in Eq.\ \eqref{eq:docc}.
      }
   \end{center}
\end{figure}

Figure \ref{fig:docc} gives the relative change of the neutrino occupation
fractions from non-degenerate FD equilibrium, i.e.,
\beq\label{eq:docc}
  \delta f(\epsilon) \equiv \frac{f(\epsilon) - \feq(\eps;0)}{\feq(\eps;0)},
\eeq
versus the dimensionless comoving energy.  The dashed lines are the
out-of-equilibrium neutrino spectra for $e$-flavor neutrinos (green) and either
$\mu$ or $\tau$-flavor (red) which are degenerate in our Boltzmann scenario.
The solid blue line is the degenerate FD equilibrium spectrum for any species
of neutrino in the \sinu scenario.  The dotted black line at zero represents a
non-degenerate spectrum at temperature \tcm.  In either scenario, there exists
an overabundance of particles at high \eps.  The positive slope of the \sinu
curve corresponds to a higher temperature \tnu compared to \tcm.  We compare
the energy scales \tnu and \tcm using the following definition
\beq\label{eq:datnu}
  \frac{\tnu}{\tcm}-1 = a\tnu - 1 \equiv \Delta(a\tnu),
\eeq
where we ascribe a unit conversion between scale factor and temperature such
that the product is dimensionless and equal to unity when $\tnu=10\,{\rm MeV}$.
At freeze-out (f.o.), we find
\beq
  \Delta(a\tnu)|_{\rm f.o.} = 2.463\times10^{-3},
\eeq
showing a slight increase in \tnu over \tcm.  Conversely, at low \eps there
exists a deficit of particles, which corresponds to a negative degeneracy
parameter
\beq
  \etanu|_{\rm f.o.} = -4.282\times10^{-3}.
\eeq

In the standard out-of-equilibrium scenario, the deficit arises from the
up-scattering of low-energy neutrinos on higher-temperature charged leptons,
given schematically in Eq.\ \eqref{eq:nue}.  The situation is similar for
\sinu, as both elastic scattering and electron-positron annihilation [Eq.\
\eqref{eq:nuepma}] add energy and a modest number of particles to the neutrino
seas. We can explain the behavior in Fig.\ \ref{fig:docc} by examining Eqs.\
\eqref{eq:dtnudt} and \eqref{eq:detanudt}.  The denominator in both equations
is positive if $\etanu=0$.  If we solve the numerators for the energy added per
particle we find
\begin{align}
  \frac{\partial\rho}{\partial n} > \frac{3\tnu n}{n,_\eta} &\implies \frac{d\tnu}{dt} > 0,\\
  \frac{\partial\rho}{\partial n} < \frac{4\rho}{3n} &\implies \frac{d\etanu}{dt} > 0,
\end{align}
where we ignore the redshift term for $d\tnu/dt$ in Eq.\ \eqref{eq:dtnudt}.
Assuming non-degenerate statistics, we calculate
\begin{align}
  \frac{3\tnu n}{n,_\eta} \simeq 3.288\tnu,\\
  \frac{4\rho}{3n} \simeq 4.202\tnu.
\end{align}
Figure [4] in Ref.\ \cite{Trans_BBN} shows that the weak-decoupling induced
distortion in the neutrino seas is located at $\eps\simeq5$ in the standard
scenario.  As this is larger than the two values above, the same energy per
particle injection in the \sinu case would imply an increase in \tnu over \tcm
and a decrease in \etanu, which is borne out by Fig.\ \ref{fig:docc}.

\begin{figure}[h]
   \begin{center}
   \includegraphics[scale=0.6]{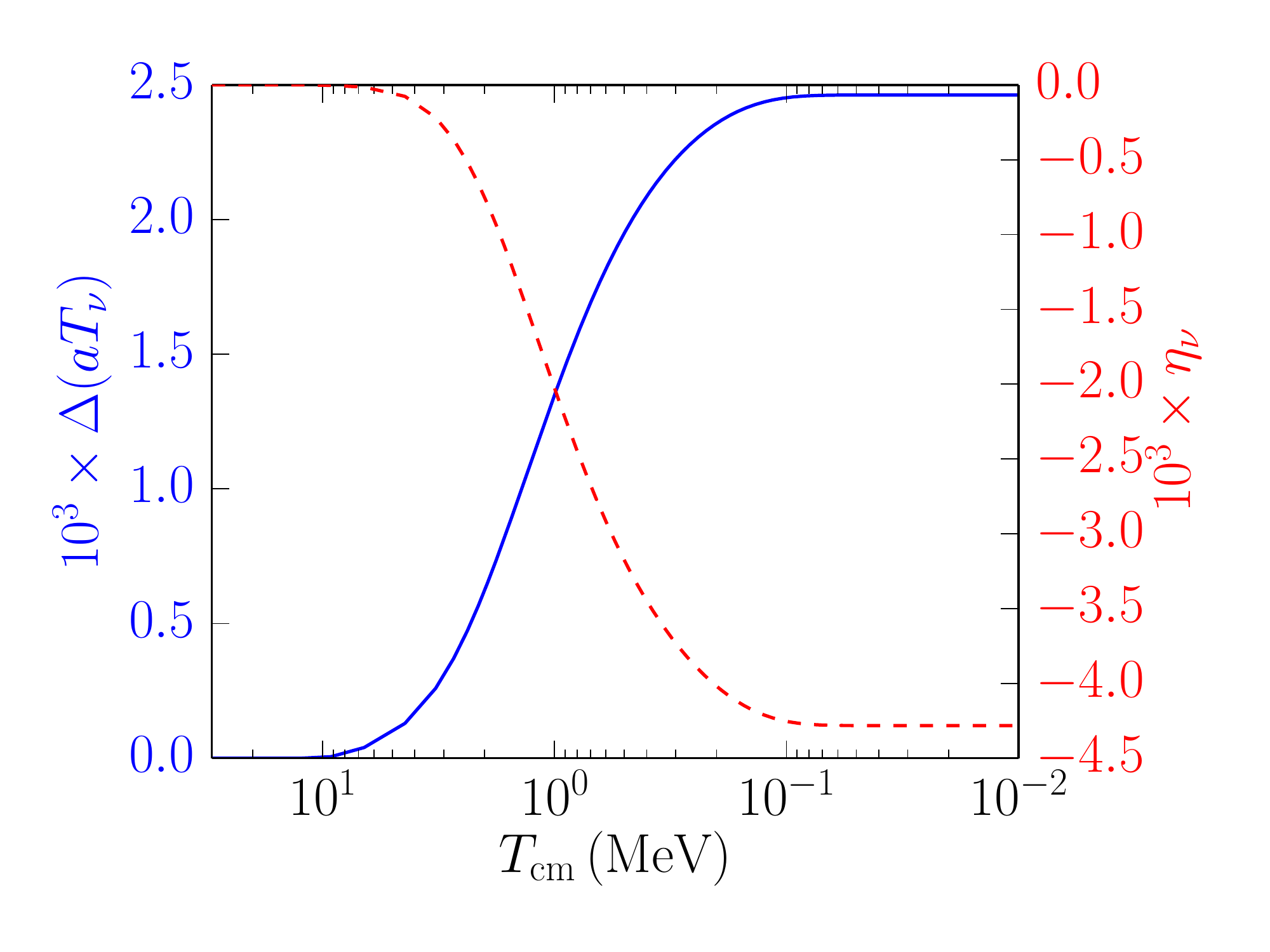}
   \caption{\label{fig:2q} Evolution of $\Delta(a\tnu)$ from Eq.\
   \eqref{eq:datnu} (solid blue curve) and \etanu (dashed red curve) plotted
   against \tcm.
      }
   \end{center}
\end{figure}

We have given the values at freeze-out for \etanu and $\tcm/\tnu$.  Figure
\ref{fig:2q} shows the evolution of $\Delta (a\tnu)$ (solid blue line) and \etanu
(dashed red line) from the initial values of zero to the freeze-out values as a
function of \tcm.  The product $a\tnu$ and the quantity \etanu can only change
during weak decoupling, i.e., when the derivatives $\partial n/\partial t|_a$
and $\partial \rho/\partial t|_a$ are non-zero.  Figure \ref{fig:2q}
illustrates the weak-decoupling epoch, starting at a few MeV, and finishing
slightly after 100 keV.

Neutrino self-interactions only indirectly influence the heat flow from the
plasma into the neutrino seas.  Figure \ref{fig:docc} shows that the occupation
numbers at freeze-out differ between the two scenarios.  This is also the case
during the entirety of weak decoupling.  Modifying the occupation numbers and
associated Pauli blocking factors will change the neutrino-charged-lepton
scattering rates through the $f^{(k)}$ factors given in Eq.\ \eqref{eq:coll}.
The \sinu spectrum lies in between the two spectra for $e$-flavor and
$\mu$-flavor in Fig.\ \ref{fig:docc}, implying a slight enhancement
(suppression) for scattering between charged leptons and $e$($\mu$)-flavor
neutrinos.  There is also an enhancement at low \eps for any flavor for a
substantially negative \etanu.  When calculating the radiation energy density
at low temperatures, we find $\neff=3.045$, a small overall increase of
$10^{-3}$ compared to the result in Sec.\ \ref{ssec:BBN_numerics}.  Although
we evolve a non-zero degeneracy parameter, our result for \neff agrees with the
$T_{\nue}=T_{\nu_{\mu,\tau}}$ model in the last row of Table [1] of Ref.\
\cite{2020arXiv200104466E}.

\subsubsection{Primordial Abundances}

The calculation for the primordial abundances is nearly identical in the
standard and \sinu scenarios.  The only difference between the two scenarios is
in the neutron-to-proton rates.  The \sinu scenario uses FD distributions with
\tnu and \etanu compared to out-of-equilibrium occupation numbers
$\{f_{\nue,\bnue}(\epsilon)\}$ for the standard scenario.  The difference
results in small relative changes for \yp and \dtoh
\begin{align}
  & \delta\yp \simeq 4\times10^{-4},\label{eq:yp_sinu}\\
%  & \delta(\dtoh) \simeq 2\times10^{-4}.\label{eq:dh_sinu}
% 25Feb2020, EG: fixed with new calculation:
  & \delta(\dtoh) \simeq 2\times10^{-4}.\label{eq:dh_sinu}
\end{align}
Both changes are well within observational uncertainties
\cite{Aver:2013ue,CP:2018d1p}.  There exists a slight increase in the ratio
$^7{\rm Li}/{\rm H}$ with a relative change of $2\times10^{-4}$.
%25Feb2020, EG: put in citation to Froustery and Pitrou (2020)
The increase in all abundances is due to an increase in the neutron-to-proton
ratio which results from a decrease in the rate $n(\nue,e^-)p$ as there exists
fewer \nue in the \sinu case \cite{2020PhRvD.101d3524F}.

\section{Dark Radiation Addition}
\label{sec:dr}

\subsection{Dark Radiation Parameter}

This section and the succeeding one explore cosmological extensions to the
\sinu scenario.  We begin by considering an addition of dark radiation
\cite{dark_rad:2000} to the universal energy density.  The dark radiation only
interacts with the other components through gravitation.  Within the
computation, only the Hubble expansion rate in Eq.\ \eqref{eq:hub} changes.  We
parameterize the dark radiation energy density by using the quantity \deltadr,
introduced in \cite{GFKP-5pts:2014mn}
\beq
  \rho_{\rm dr} = \frac{7\pi^2}{120}\tcm^4\deltadr.
\eeq
After electron-positron annihilation terminates, we can calculate the change to
\neff by adding $\rho_{\rm dr}$ to \rhorad in Eq.\ \eqref{eq:neff}
\beq\label{eq:dneff_dr}
  \Delta\neff^{\rm (dr)} = \left(\frac{11}{4}\right)^{4/3}
  \left(\frac{\tcm}{T}\right)_{\rm f.o.}^{4/3}\deltadr.
\eeq
Weak decoupling and QED corrections to the plasma equation of state modify the
transfer of entropy from the charged leptons to the photons.  The result is a
freeze-out ratio $\tcm/T$ larger than $(4/11)^{1/3}$, causing $\Delta\neff^{\rm
(dr)}$ to be larger than \deltadr.  The increase is less than 1\%, so
$\Delta\neff^{\rm (dr)}\approx\deltadr$ to good approximation.

We do a parameter-space scan over \deltadr to calculate the changes to the
neutrino spectra and primordial abundances in the \sinu scenario.

\subsection{Results}

\subsubsection{Neutrino Spectra}

By construction, the dark radiation component has no direct effect on neutrino
scattering and weak decoupling.  When adding extra radiation energy density to
the universe, we expect an earlier epoch of weak decoupling precipitated by the
more rapid expansion rate.  This conclusion is identical for either the
standard or \sinu scenarios.  The final freeze-out spectra look similar
to the blue curve in Fig.\ \ref{fig:docc}, and the evolution of \tnu and \etanu
are qualitatively the same for the scenarios with and without dark radiation.

\begin{figure}[h]
   \begin{center}
   \includegraphics[scale=0.6]{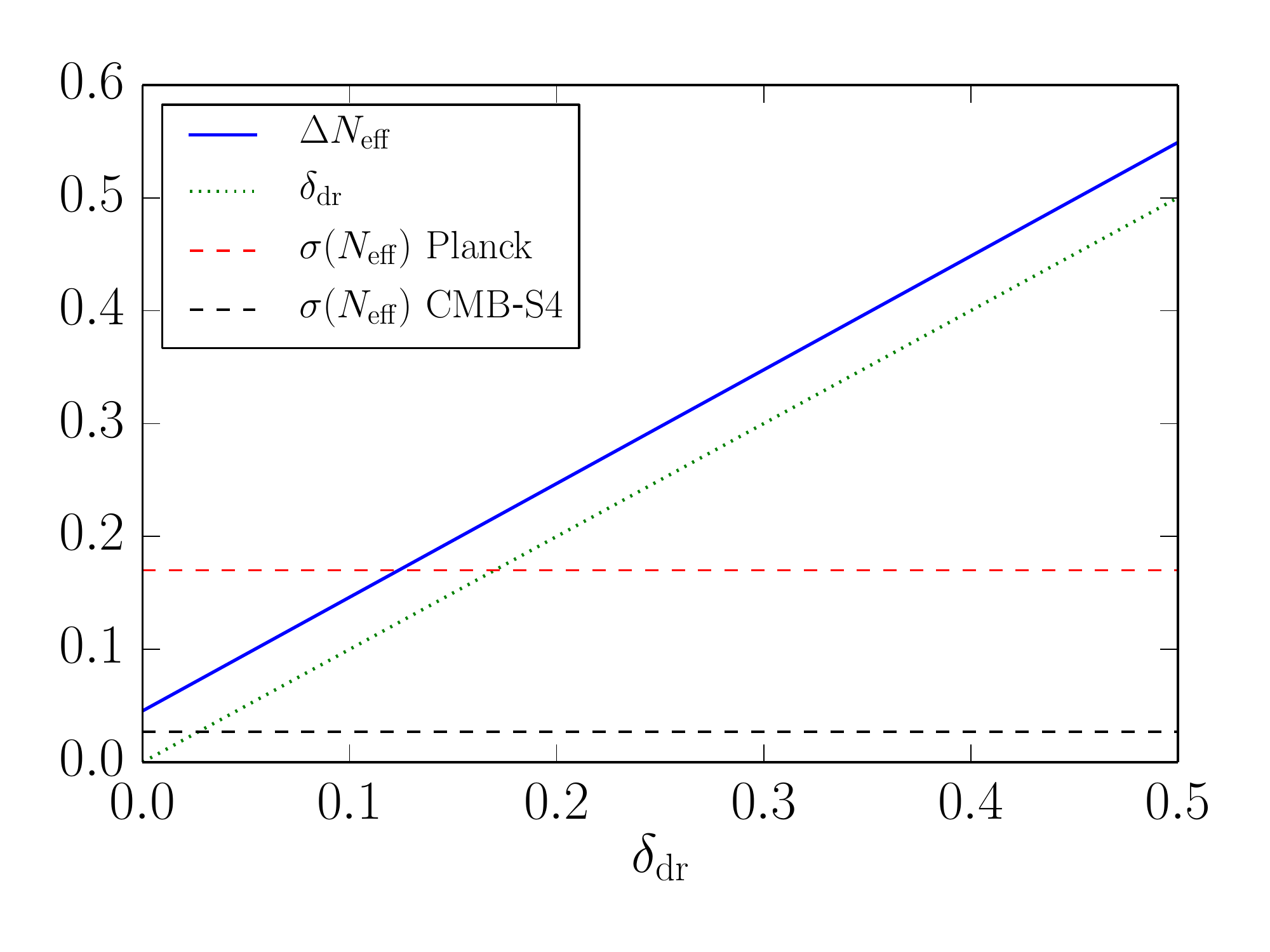}
   \caption{\label{fig:dneff} Change in \neff plotted against \deltadr.
      }
   \end{center}
\end{figure}

In Fig.\ \ref{fig:dneff} we plot four quantities related to \neff against the
parameter \deltadr.  The solid blue curve gives $\Delta\neff\equiv\neff-3$ and
includes neutrino transport, QED corrections, and dark radiation in the \sinu
scenario.  The dashed green curve is \deltadr and close to the change in \neff
from the addition of the dark radiation.  The horizontal lines give the
$1\sigma$ uncertainty in \neff from Planck (dashed red line)
\cite{2018arXiv180706209P} and a projection from Cosmic Microwave Background (CMB) Stage IV (dashed black
line) \cite{cmb_s4_white_paper}.

The blue and green lines appear to run parallel to one another in Fig.\
\ref{fig:dneff}, suggesting that changes to \neff from neutrino transport/QED
corrections and dark radiation add linearly.  This conclusion is indeed the
case to good precision and is also true in the standard scenario.  In models
with dark radiation, the non-linear contribution to $\Delta\neff$ comes from a
change in the expansion rate which leads to an earlier epoch of weak decoupling
and subsequently lowers the amount of heat flow from the plasma into the
neutrino seas.  This effect changes both the energy density in the neutrino
seas and also the ratio $\tcm/T|_{\rm f.o.}$ in Eq.\ \eqref{eq:dneff_dr}.  We
have verified this behavior by comparing $\neff(\deltadr=0.5)$ to
$\neff(\deltadr=1.0)$ in the following manner
\beq\label{eq:non_lin_dr}
  2 - \frac{\neff(\deltadr=1.0) - \neff(\deltadr=0)}
  {\neff(\deltadr=0.5) - \neff(\deltadr=0)} \simeq 9\times10^{-5},
\eeq
indicating that the degree of non-linearity is small and the blue and green
curves in Fig.\ \ref{fig:dneff} are parallel to high precision. The
non-linearity grows with increasing \deltadr but is not a factor for modest
values of \neff.  We emphasize this analysis is for the \sinu scenario but is
directly applicable to the standard scenario.

\subsubsection{Primordial Abundances}

Increasing \deltadr precipitates an earlier epoch of weak freeze-out between
the neutrinos and baryons, and results in a higher neutron-to-proton ratio
during nuclear freeze-out.  Both \yp and \dtoh increase in the presence of an
over-abundance of neutrons.  We direct the reader to Refs.\
\cite{2002PhRvD..65b3511H,Hamann:2008bc} for more information on extra
radiation energy density in BBN, and in particular Ref.\
\cite{2004NJPh....6..117K} for quantitative scaling forms of the abundances
with $\Delta\neff$.  Transport and self-interactions do little to modify the
scaling forms and instead renormalize the values when $\deltadr=0$.  We give
specific values of \yp and \dtoh with relative changes from the standard
baseline for the case $\deltadr=0.4009$
\beq\label{eq:dr_abunds}
\begin{array}{cccr}
  \yp = 0.2532&\implies&\delta\yp = 2.161\%&\quad(\deltadr=0.4009),\\
  \dtoh = 2.769\times10^{-5}&\implies&\delta(\dtoh) = 5.385\%&\quad(\deltadr=0.4009).
\end{array}
\eeq

We highlight that Eq.\ \eqref{eq:dr_abunds} shows that deuterium, not helium,
has a larger relative change with the addition of dark radiation.  In fact,
this is a general trend in BBN and not specific to the value of \deltadr taken
in Eq.\ \eqref{eq:dr_abunds}.  Figure 10 of Ref.\ \cite{2004NJPh....6..117K}
shows the hyper-sensitivity of \dtoh to extra energy density.  The reason for
the result in Eq.\ \eqref{eq:dr_abunds} is the dark radiation precipitates an
earlier freeze-out of nuclear reaction rates.  Given the precision in measuring
\dtoh, any BBN scenario which includes a dark radiation component must consider
the impact on deuterium production.

\section{Lepton Asymmetric Initial Conditions}
\label{sec:w_deg}

We give a second example of a one-parameter extension to the \sinu scenario by
introducing an asymmetry between numbers of neutrinos and anti-neutrinos with the
following quantity \cite{lep_trans}
\beq\label{eq:lnustar}
  L_{\nu_i}^{\star} \equiv \frac{n_{\nu_i}-n_{\overline{\nu}_i}}{\frac{2\zeta(3)}{\pi^2}\tcm^3},
  \quad i=e,\mu,\tau,
\eeq
where $\zeta(3)=1.202$ is the Riemann Zeta function of argument 3.  The above
definition is identical with the standard definition of lepton number
\beq\label{eq:lep_num}
  L_{\nu_i} = \frac{n_{\nu_i}-n_{\overline{\nu}_i}}{n_\gamma},
\eeq
at high temperature where $\tcm=T$.  We use $L_{\nu_i}^{\star}$ as it is a
comoving invariant throughout electron-positron annihilation and label it the
comoving lepton number to distinguish it from Eq.\ \eqref{eq:lep_num}.  If
self-interactions violate flavor conservation, than we would expect the
comoving lepton number to be the same in all three flavors.  Additionally, if
self-interactions violate lepton number, we would expect $L_{\nu_i}^{\star}=0$
for all flavors at the commencement of BBN.  We will consider the \sinu
scenario of flavor-violation and lepton-number conservation in this section.
Therefore, we only use a single quantity \lnustar to describe the asymmetry for
any flavor.

\subsection{Differences in Implementation with Symmetric Scenario}

Neutrinos and anti-neutrinos will evolve uniquely in the \sinu scenario with
asymmetric initial conditions.  In general, we need temperature and
degeneracy quantities for the anti-neutrinos, namely, $T_{\bnu}$ and
$\eta_{\bnu}$. 
The Lagrangian in Eq.\ \eqref{eq:l_int} allows for $t$-channel scattering between neutrinos and anti-neutrinos, so the interaction
\beq\label{eq:nu_bnu}
  \nu + \bnu\leftrightarrow\nu+\bnu,
\eeq
holds the two seas in thermal equilibrium, implying $\tnu=T_{\bnu}$.  Equation \eqref{eq:nu_bnu}, however, does not hold the two seas in chemical equilibrium.  As a result, our model of \sinu with lepton degeneracy only requires the introduction of one new variable, namely $\eta_{\bnu}$.
The set of variables $\{\tnu,\etanu,\eta_{\bnu}\}$ evolve in accordance with the definitions of
energy and number density for neutrinos and anti-neutrinos.  The expressions for $n_{\nu}(\tnu,\etanu)$ and $n_{\bnu}(\tnu,\eta_{\bnu})$ are identical to Eq.\
\eqref{eq:nrho}, except now $N_\nu=3$ for three flavors of
neutrinos/anti-neutrinos. $\rho_\nu$ and $\rho_{\bnu}$ follow from analogy.

The asymmetry in the system implies there will be heat flow between the
neutrinos and anti-neutrinos.  Although the weak interaction can precipitate such a heat flow, the $t$-channel \sinu interaction will be much more efficient and will thermally equilibrate the two seas instantaneously.  To account
for this fact, we need to supplement the weak collision terms
 with an equilibration term $d\mathcal{E}/dt$
\begin{align}
  \frac{\partial\rho_\nu}{\partial t}\biggr|_a &= \frac{\tnu^4}{2\pi^2}\sum\limits_{i=1}^3
  \int d\eps\,\eps^3\{\cnue[\feqene] + \cepma[\feqene]\} + \frac{d\mathcal{E}}{dt},\\
  \frac{\partial\rho_{\bnu}}{\partial t}\biggr|_a &= \frac{\tnu^4}{2\pi^2}\sum\limits_{i=1}^3
  \int d\eps\,\eps^3\{\cnue[f^{\rm (eq)}(\eps;\eta_{\bnu})] + \cepma[f^{\rm (eq)}(\eps;\eta_{\bnu})]\} - \frac{d\mathcal{E}}{dt}.
\end{align}
The equilibration term is a weighted average of the energy and number flows due to the weak interaction.  If we group the weak-collision terms into $d\rho_\nu/dt|_{\rm w}$ and $d\rho_{\bnu}/dt|_{\rm w}$, we find
\begin{align}
  \frac{d\mathcal{E}}{dt} &=
  [4n_{\nu,\eta}n_{\bnu,\eta}(\rho_\nu+\rho_{\bnu})-9\tnu(n_\nu^2n_{\bnu,\eta} + n_{\bnu}^2n_{\nu,\eta})]^{-1}\nonumber\\
  &\times\left\{
  3\tnu[4(n_\nu\rho_{\bnu}n_{\bnu,\eta} - n_{\bnu}\rho_{\nu}n_{\nu,\eta})
  +9\tnu n_\nu n_{\bnu}(n_\nu-n_{\bnu})]\frac{\partial n_{\nu}}{\partial t}\biggr|_a\right.\nonumber\\
  &\left.+ 4n_{\nu,\eta}n_{\bnu,\eta}\left(\rho_\nu\frac{\partial\rho_{\bnu}}{\partial t}\biggr|_{\rm w}
  -\rho_\nu\frac{\partial \rho_{\bnu}}{\partial t}\biggr|_{\rm w}\right)
  +9\tnu\left(n_{\bnu}^2n_{\nu,\eta}\frac{\partial \rho_{\nu}}{\partial t}\biggr|_{\rm w}
  - n_\nu^2n_{\bnu,\eta}\frac{\partial \rho_{\bnu}}{\partial t}\biggr|_{\rm w}\right)\right\}\label{eq:equil_term}
\end{align}
In writing Eq.\ \eqref{eq:equil_term}, we have set $\partial n_{\bnu}/\partial t|_a = \partial n_\nu/\partial t|_a$ as the weak-collision terms conserve lepton number.  Those number-flow terms are analogous to Eq.\ \eqref{eq:sinu_num_flow}.  The derivatives for \tnu, \etanu, and $\eta_{\bnu}$ follow from Eqs.\ \eqref{eq:dtnudt} and \eqref{eq:detanudt}.

Initially at high temperature, neutrinos and anti-neutrinos are in thermal
equilibrium with the plasma and $\tnu=T$.  In addition, the weak interaction ensures neutrinos
and anti-neutrinos are in chemical equilibrium and have degeneracy parameters
equal in magnitude and opposite in sign.  The equation
\beq
  \lnustar = \frac{1}{12\zeta(3)}(\pi^2\xi + \xi^3)
\eeq
relates the comoving lepton number to the degeneracy parameter
\cite{1990eaun.book.....K}.  We begin the calculation by setting $\etanu=\xi$
and $\eta_{\bnu}=-\xi$.

\subsection{Results}

\subsubsection{Neutrino Spectra}

\begin{figure}[h]
   \begin{center}
   \includegraphics[scale=0.6]{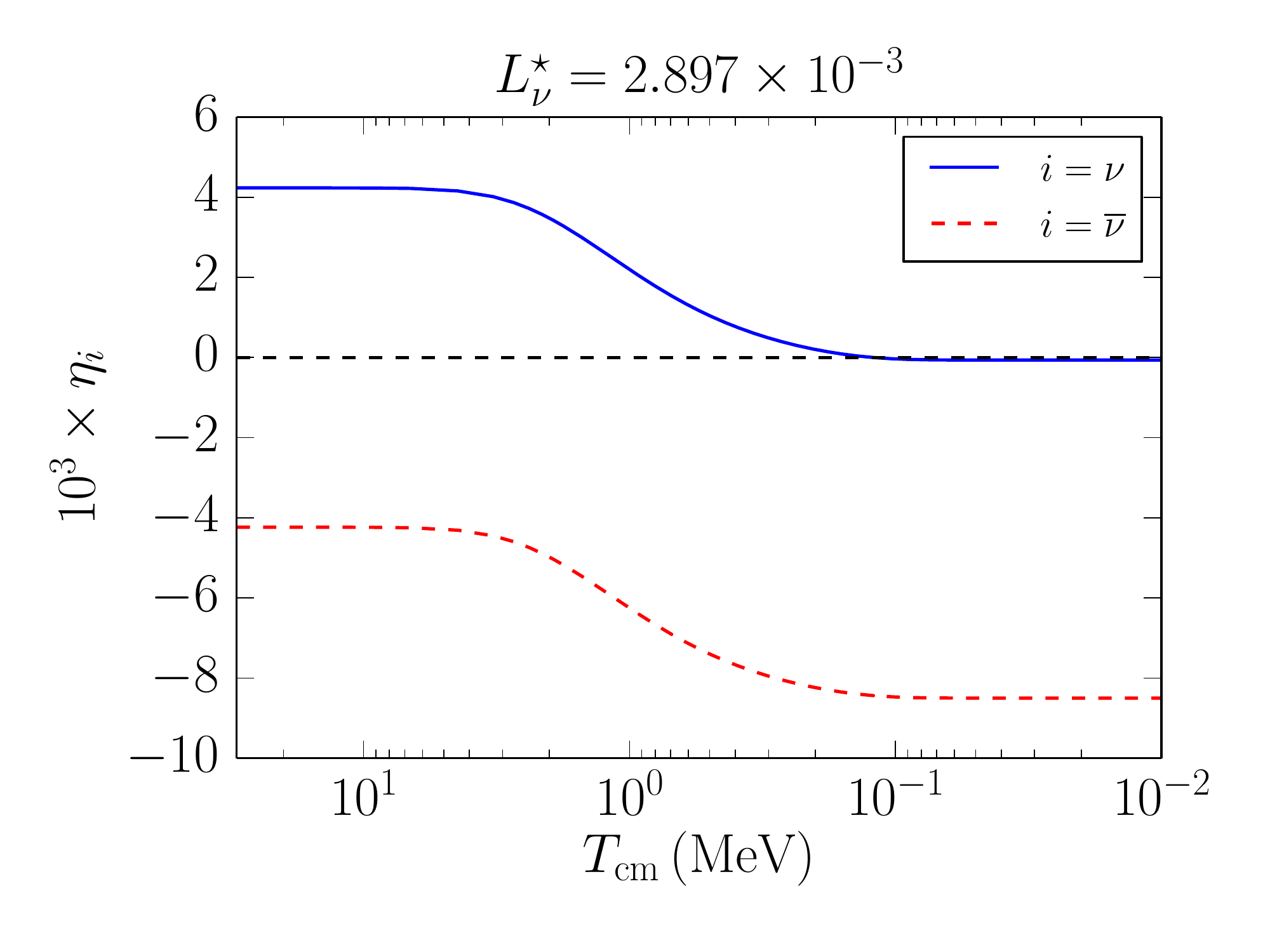}
   \caption{\label{fig:deg_comp_eta} Evolution of $\eta_i, i=\nu,\bnu$
   as a function of \tcm.  The comoving lepton number is $\lnustar=2.897\times10^{-3}$.
      }
   \end{center}
\end{figure}

We give an example of an asymmetric \sinu calculation with
$\lnustar=2.897\times10^{-3}$ in Fig.\ \ref{fig:deg_comp_eta}.  The
two curves of Fig.\ \ref{fig:deg_comp_eta} show the evolution of the
degeneracy parameters for neutrinos (solid blue curve) and anti-neutrinos
(solid red curve) as a function of \tcm.  These two curves evolve
similarly albeit with a noticeable offset due to the initial asymmetry.
Figure \ref{fig:deg_comp_eta} clearly shows the composite neutrino/anti-neutrino system is not
in equilibrium as $\etanu\ne-\eta_{\bnu}$ once weak decoupling begins.
We do not show the evolution of \tnu as its behavior is similar to that of \tnu in the lepton-symmetric case in Fig.\ \ref{fig:2q}.
The paucity of anti-neutrinos leads to a suppression of the Pauli blocking factors in the weak collision terms, and therefore an increase in the scattering rates of charged leptons with anti-neutrinos.  Conversely, the overabundance of neutrinos enhances the Pauli blocking factors and decreases those scattering rates.  The two effects together imply $d\mathcal{E}/dt>0$.  Overall, the increase in the anti-neutrino scattering rates and the decrease in the neutrino rates nearly cancel and the end result is the evolution of \tnu in the lepton-asymmetric case differs only slightly from the lepton-symmetric case.
The lepton-asymmetric case does induce a larger heat flow than the lepton-symmetric case, but the difference is below the resolution of the plot for \tnu in Fig.\ \ref{fig:2q}.

We have shown results for the neutrino parameters in the case $\lnustar>0$.  If
$\lnustar<0$, then we are in the opposite $CP$ system and the results
corresponding to Fig.\ \ref{fig:deg_comp_eta} would be $CP$ conjugates.  The
quantity \neff does change in an asymmetric scenario as this system has a
larger energy density.  However, \neff only has sensitivity to the magnitude of
\lnustar.  Both the larger energy density from degenerate FD statistics and
neutrino energy transport are symmetrical for $\pm\lnustar$.  Therefore, \neff
is immune to the sign of \lnustar.

\subsubsection{Primordial Abundances}

\begin{figure}[h]
   \begin{center}
   \includegraphics[scale=0.6]{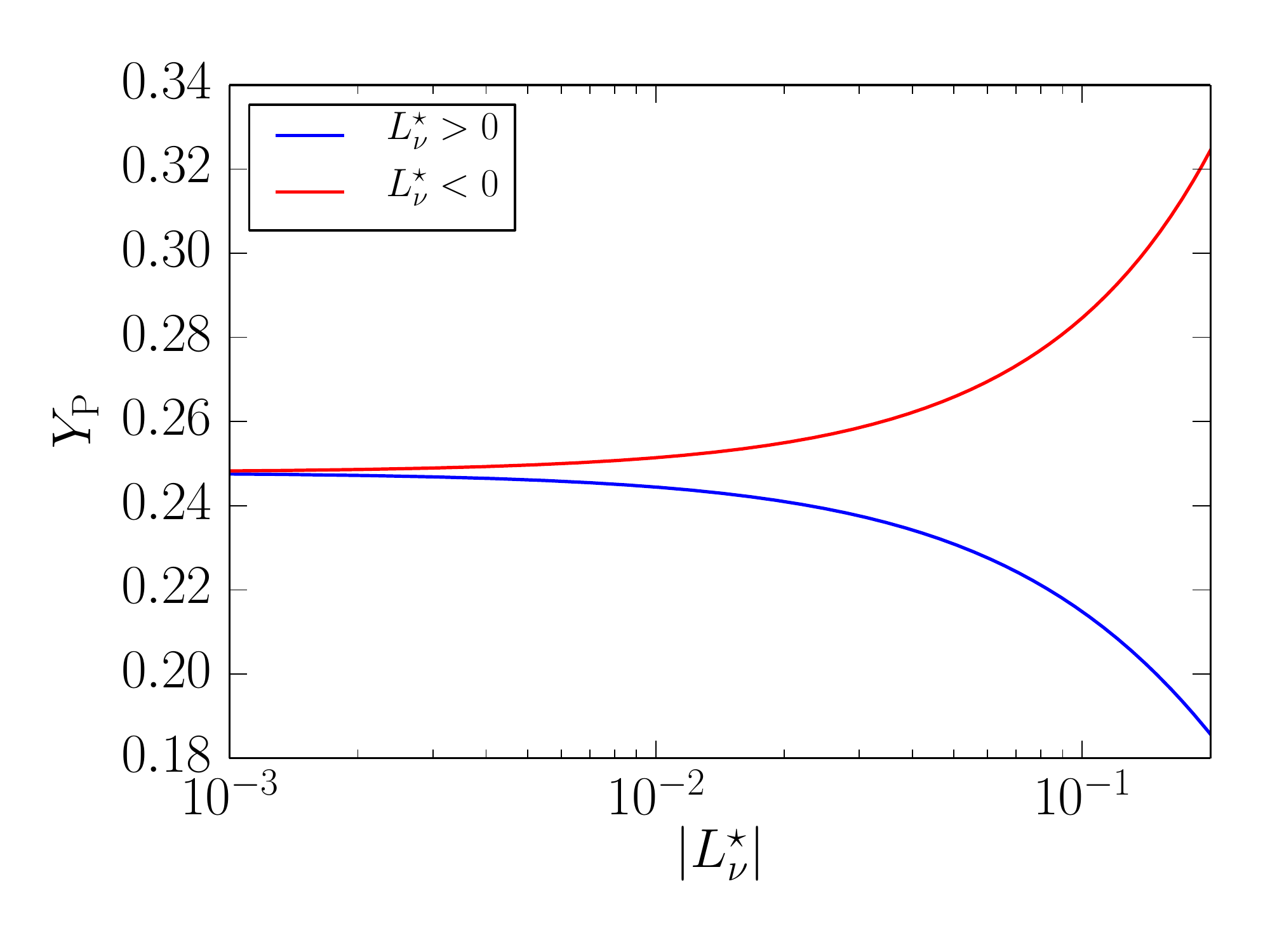}
   \caption{\label{fig:deg_yp} \yp plotted against $|\lnustar|$ [Eq.\
   \eqref{eq:lnustar}] in the \sinu scenario.}
   \end{center}
\end{figure}

As opposed to \neff, the primordial abundances are indeed sensitive to the sign
of \lnustar.  The neutrino spectra are inputs into the neutron-to-proton rates
and therefore change the evolution of the neutron-to-proton ratio.  We execute
a parameter space scan over \lnustar to examine how the lepton asymmetry
affects the primordial abundances in the \sinu scenario.  Figure
\ref{fig:deg_yp} shows \yp as a function of $|\lnustar|$, where the blue curve
gives \yp for $\lnustar>0$ and the red curve gives \yp for $\lnustar<0$.  We
only show \yp in Fig.\ \ref{fig:deg_yp}, but the other abundances have similar
shapes as seen in Fig.\ [14] in Ref.\ \cite{lep_trans}.  In particular, \dtoh
follows qualitatively the same shape as the curves in Fig.\ \ref{fig:deg_yp}
and has a relative change of $\sim1/2$ as compared to \yp at all values of
\lnustar.

Before we conclude, we provide an example model with specific values.  If we
desire to decrease the helium mass fraction by 10\%, we would need a positive
comoving lepton number of $\lnustar\simeq0.073$.  A 10\% change in \yp would be
difficult to reconcile with current observational bounds \cite{Aver:2013ue}.
The extra energy density from this value of comoving lepton number translates
to $\Delta\neff=0.06$, which is consistent with Ref.\
\cite{2018arXiv180706209P}.  We conclude that \yp has the potential to be a
better probe for \lnustar than \neff.  Our conclusion is for the \sinu scenario
and is directly applicable to the standard scenario extension with lepton
asymmetry.

\section{Conclusions}
\label{sec:concl} 

\subsection{Summary}
\label{ssec:sum}

Self-interacting neutrinos constitute a model of Beyond Standard Model (BSM) physics with implications for BBN.
Neutrinos go out of thermal and chemical equilibrium during BBN in the standard
scenario.  Figure \ref{fig:docc} shows that these deviations from equilibrium
are small, although potentially detectable in future measurements of \neff
\cite{cmb_s4_white_paper}.  Self-interacting neutrinos have little influence on
the magnitude of the deviations, and instead reshuffle the neutrino energy
among the flavors while preserving the general shape of the spectra.  Figure
\ref{fig:2q} shows that self-interactions modify the spectral parameters by a
few parts in $10^3$.  Although the physics governing neutrino interactions is
different, the standard and \sinu scenarios do not predict any substantial
differences for cosmological observables/parameters at present or future
precision [see Eqs.\ \eqref{eq:yp_sinu} and \eqref{eq:dh_sinu}].

We have extended our BSM model to include self-interactions and dark radiation.
With the inclusion of extra radiation energy density, weak decoupling
terminates at an earlier epoch and less heat flows from the electromagnetic
plasma to the neutrino seas.  Equation \eqref{eq:non_lin_dr} shows the change
in neutrino energy density caused by the earlier freeze-out is small.  As a
result, self-interactions perturb the neutrino seas in much the same manner as
they did without the addition of the dark radiation.  Figure \ref{fig:dneff}
shows how \neff changes with the addition of dark radiation, which is nearly an
incoherent sum of the extra energy density and the effects from weak decoupling.
Although Fig.\ \ref{fig:dneff} is for the \sinu scenario, \neff behaves in
quantitatively the same manner for the one-parameter extension to the standard
scenario.

Finally, we extended the \sinu scenario to include lepton asymmetry.  In the
class of models we explore, the self-interactions always maintain
flavor-identical distributions among the neutrino and anti-neutrino seas.  The
flavor-independence of the self-interactions results in the lepton numbers
necessarily being identical to one another for the three flavors.  The
difference between this scenario and the ones with symmetric initial conditions
is that the neutrino and anti-neutrino components can have different spectra.
Figure \ref{fig:deg_comp_eta} shows different evolutions for the
degeneracy parameters in a \sinu scenario where
$\lnustar=2.897\times10^{-3}$.  Although the 
degeneracy parameters undergo different evolutions compared to the
symmetric scenario, the net change from the initial starting point is still a
few parts in $10^3$. As a result, predictions on cosmological
parameters/observables mainly scale with the presence of a non-zero \lnustar,
and self-interactions produce higher-order effects.  Figure \ref{fig:deg_yp}
gives the primordial mass fraction of $^4{\rm He}$ as a function of \lnustar in
the \sinu scenario.  Lepton degeneracy sources an increasing (decreasing) \yp
with decreasing (increasing) \lnustar while self-interactions simply translate the
curves to slightly higher values.

\subsection{Discussion}
\label{ssec:disc}

The motivation for this work was to determine the dynamics of weak decoupling
and primordial nucleosynthesis in the presence of \sinu.  Our impetus for this
decision is two-fold.  First, the neutrinos experience kinetic evolution
during this epoch, as scattering rates are neither large enough to maintain
equilibrium nor small enough to preclude thermal contact.  If new neutrino
physics manifests during BBN, standard-model physics will not be able to erase
these signals.  Second, cosmological parameters/observables directly influenced
by the physics of the BBN epoch have already been measured.  Advances in
precision on $\{\neff,\yp,\dtoh\}$ will either place stronger constraints or
reveal signatures of new physics. Coupled together, the kinetic regime and the
ability to make measurements which provide diagnostics of the physics operating during this epoch provide researchers with a tool to
explore new physical theories.

Our intent for this work was to quantify how self-interactions, and solely
self-interactions, modify the neutrino spectra and concomitant light-element
synthesis.  Section \ref{sec:no_deg} presents the results for the \sinu
scenario which amount to less than a 1\% change from the standard BBN scenario.
The same level of relative changes between the standard and \sinu scenarios
occur when we add dark radiation or lepton asymmetry to BBN.  We conclude that
BBN does not provide any additional constraints on a scenario with only
self-interacting neutrinos.

We chose a limited class of \sinu models for this work.  Namely, we operate in
the following \sinu framework: (1) the interaction matrix, $g_{ij}$ in Eq.\
\eqref{eq:g_matrix}, is agnostic to any particular flavor; (2) the
self-interactions are $CP$-symmetric; (3) $\varphi$ is a real scalar; (4) the mass of
$\varphi$ is large compared to BBN energy scales; and finally (5), the neutrinos
stay self-coupled throughout BBN.  We contemplate the repercussions if we relax
each one of these limitations in turn.

Our choice for $g_{ij}$ forced all flavors of neutrinos to have the same
distribution.  Had we set the off-diagonal terms of $g_{ij}$ to zero, then we
would have needed separate parameters $T_{\nu_i}$ and $\eta_{\nu_i}$ for each
flavor.  Given how little self-interactions changes cosmological
parameters/observables, we do not expect this class of models to alter our
conclusions significantly.  Table [1] in Ref.\ \cite{2020arXiv200104466E}
appears to bear out this conclusion for \neff.  Moreover, had we introduced a
flavor asymmetry in the interaction matrix, i.e., $g_{ee}\ne g_{\mu\mu}$, then
we might expect some flavors of neutrinos to stay in self-equilibrium and
others to evolve in an out-of-equilibrium fashion.  Figure \ref{fig:docc} shows
there is little difference between these two kinds of distributions, so again
we expect little effect on parameters.  What may be different would be the case
where there exists lepton asymmetries.  If the self-interacting Lagrangian is
indeed flavor biased, then lepton numbers can be different for different
flavors, and oscillations would play a dominant effect during BBN
\cite{2002NuPhB.632..363D,2012PhRvD..86b3517C,2016PhRvD..94h3505J}.
Furthermore, the flavor-asymmetric scenario could have severe repercussions on
neutrino free-streaming during atomic recombination, if the neutrino gas is
only partially self-coupled.  Such a scenario requires a calculation with
vacuum neutrino oscillations, but we can give a qualitative description here.
In the standard cosmology, free-streaming neutrinos cause a phase shift in the
CMB acoustic peaks \cite{2004PhRvD..69h3002B,2015PhRvL.115i1301F}.  The \sinu
gas eliminates that phase shift and utilizes extra dark radiation to remedy
this particular temperature power-spectrum defect.  If not all the neutrinos
were to be self-coupled, then the free-streaming neutrino component would
induce a small phase shift, and the amount of dark radiation would necessarily
be smaller.  At the current epoch, the effect would be to increase the Hubble
parameter only slightly.

In addition to all flavors being self-interacting, \sinu scenarios which remedy
the Hubble parameter tension also stipulate the anti-neutrinos to be
self-interacting.  If the interacting Lagrangian in Eq.\ \eqref{eq:l_int} is
not $CP$-symmetric, then models which use \sinu to resolve cosmological
tensions will also confront the same problems as the flavor-asymmetric models
discussed above.  From the perspective of BBN, if anti-neutrinos do not remain
in a self-equilibrium, then the anti-neutrino spectra would follow the dashed
lines in Fig.\ \ref{fig:docc} while the neutrinos followed the solid line.
Again, we do not expect the asymmetry to cause significant changes in
cosmological parameters related to BBN.  With the introduction of a lepton
number, we still would expect equal lepton numbers in all flavors as
self-interactions will change the flavor content in the neutrino sector and
standard model interactions will then equilibrate the numbers of neutrinos and
anti-neutrinos such that all have identical chemical potentials.

In our analyses of neutrino-anti-neutrino asymmetry, we have always made the
assumption that the net lepton number (i.e., the sum of the three flavor lepton
numbers) cannot change.  For the Lagrangian in Eq.\ \eqref{eq:l_int}, this is
certainly true.
We could alter the Lagrangian to include right-handed neutrinos, with an operator such as $\overline{\nu_R}\nu_L\varphi$.  For Majorana character, left-handed states could undergo spin-flips to the right-handed states at the same rate as the processes which equilibrate the neutrino/anti-neutrino seas.  A lepton asymmetry cannot persist in such a scenario.  If neutrinos are Dirac fermions, then the spin-flip populates the opposite helicity states.  These states are inactive under the weak interaction, but would still have ramifications on BBN as they contribute to the radiation energy density and would raise \neff.  The spin-flip interactions would have to freeze-out well before BBN \cite{cmbs4_science_book} for a \sinu model with right-handed neutrinos to be viable.  As a corollary, BBN would proceed without \sinu interactions but with the addition of dark radiation.
Alternatively, if we were to consider vector mediators, then lepton number
would not be conserved and we would expect a symmetric system.  We do not
anticipate any further implications for either BBN or photon decoupling if the
mediator is a vector.
Another possibility in the class of \sinu models is for a pseudoscalar mediator.  In this scenario, neutrinos may recouple after weak decoupling \cite{2014JCAP...07..046A}.  Some late portion of recombination could transpire with equilibrium neutrino spectra, but BBN would occur with out-of-equilibrium distributions identical to the standard scenario.

Returning to the scalar-mediator scenario, we only considered the case when the
mass of the mediator is much larger than BBN energy scales.  If this were not
the case, then it is possible for the mediator to be produced on shell, through
reactions like $\nu+\nu\rightarrow \phi+\phi$, during BBN.  As we take the
self-interactions strong enough to ensure neutrino self-equilibrium throughout
BBN, light mediators will continue to remain in equilibrium with the neutrino
seas until the neutrinos decouple from themselves at late times.  There are
three ramifications of this.  First, the $\varphi$-sea acts as a heat bath for
neutrinos without conserving number.  With a heat bath and strong coupling, the
neutrino seas are in thermal and chemical equilibrium with temperature \tnu and
vanishing \etanu.  The evolution of \tnu and \etanu in Eqs.\ \eqref{eq:dtnudt}
and \eqref{eq:detanudt} no longer apply, and we would have to construct an
analog of Eq.\ \eqref{eq:dtempdt} for the $\nu$-$\varphi$ system.  Second, if a
$\varphi$-sea persists to times after weak decoupling, then there will be extra
energy density in the $\nu$-$\varphi$ system which will become only neutrino
and therefore increase the radiation energy density and \neff.  This extra
energy density is not dark radiation and will have ramifications on temperature
power spectra \cite{2019PhRvL.123s1102B}.  Lastly, the energy in the
$\varphi$-sea will heat the neutrinos to higher temperatures.  If this occurs
prior to helium formation, the higher neutrino temperature will cause a lower
neutron-to-proton ratio and a smaller amount of helium.

Our final retrospection on the \sinu framework presented in this paper, is if
we remove the stipulation that the neutrinos stayed in self-equilibrium
throughout BBN.  This is a class of models where $G_{\rm eff}\sim G_F$ and
would not be able to alleviate the Hubble tension.  As the self-interactions
only reshuffle neutrino energy density, we would expect changes to BBN
observables to be somewhere between the standard and full \sinu scenarios.  In
this case, we would not expect significant changes in $\{\neff,\yp,\dtoh\}$.
The more interesting scenario would be for the case of mediators with masses
low enough to have non-zero densities during BBN.  If the self-interactions
freeze-out prior to or during BBN, the population of $\varphi$ may still be
present and decay into high-energy neutrinos.  Such a decay is necessarily an
out-of-equilibrium process and would require a transport calculation.  If this
decay occurs during weak decoupling, there exists the possibility that energy
and entropy flow from the neutrino seas into the electromagnetic plasma,
opposite of the direction in standard weak decoupling.  The abundances,
especially \dtoh, are sensitive to this entropy flow.  Given the precision on
measuring \dtoh, BBN has the potential of putting stricter limits for the
``weak-self-interacting'' case than the stronger case considered in this work.

There can be other, non-cosmology constraints on some self-interacting sterile neutrino models.
Such models with low enough scalar mass $m_\varphi$ and utilizing Majorana neutrinos can produce a unique neutrino-less double beta decay signature. This is where the double beta decay nuclear mass difference is shared among two electrons and an undetected scalar \cite{1987PhRvL..59.2020E}. This distinctive phase space signature has allowed Majoron-like models \cite{1981PhLB...99..411G,1981NuPhB.193..297G,1985PhLB..159...57G} with effective couplings $g > {10}^{-3}$ \cite{2019PhRvD..99i6005B} to be constrained.  Future tonne-scale double beta decay detectors will have higher sensitivity and therefore likely will be able to extend these constraints. Finally, the $Z^0$-width measurement obviously provides a stringent limit on the triplet versions but not the singlet versions of the Majoron-like models discussed above. There can be other such constraints, for example on four-neutrino couplings \cite{Belotsky:2001fb}.

Neutrino physics can play a significant, even dominant role in the dynamics, evolution, and nucleosynthesis in some compact object environments. These include, for example, core collapse supernovae and binary neutron star mergers. Not surprisingly, BSM neutrino self-interactions may have dramatic effects in these venues. These effects could lead to constraints which can complement those derived from the early universe considerations discussed in this work. However, the origin of some of the potentially constrainable compact object effects is rooted in the same self-equilibration and flavor changing physics that affects weak decoupling and BBN in the early universe. When it comes to neutrino self-interactions and flavor, the key difference between compact objects and the early universe is the very low entropy per baryon in the former venue. For example, the collapsing cores of supernovae are characterized by low entropy and highly degenerate electron lepton number carried by seas of electrons and electron neutrinos. When the neutrinos become trapped by coherent neutral current scattering on heavy nuclei, self-interactions can do two things: (1) effect rapid self-equilibration among the neutrinos, very much like the self-equilibration in the early universe discussed above; and (2) rapidly re-distribute the degenerate electron lepton number among all three neutrino flavors. Electron neutrinos can be converted to $\mu$ and $\tau$ species, opening phase space for electron capture, $e^-+p \rightarrow n+\nu_e$. Very quickly the electron lepton number residing in the degenerate sea of electrons is redistributed and entropy is generated. This can result in dramatic alteration of the physics of collapse \cite{Konoplich:1988mj,1988ApJ...332..826F}. Moreover, after core bounce, the neutrinos can be \lq\lq cooled\rq\rq\ by adiabatic expansion of the strongly self-coupled neutrino gas. This could lead to a constraint if neutrino energies are reduced below the thresholds for the IMB \cite{1987PhRvL..58.1494B} and Kamiokande II \cite{1987PhRvL..58.1490H} experiments which detected the neutrino burst from supernova SN1987a \cite{2019arXiv191209115S}.

In closing, we mention the importance of deuterium in BBN
constraints.  Although the topic of this work is \sinu, our analysis applies to
other BSM scenarios.  Cosmological models which posit extra radiation energy
density during BBN \cite{2014PhRvL.112e1302W,2014PhRvD..90f3009P,2019arXiv190610136G,2019arXiv191101176K} necessarily change the abundances.  In particular, helium and deuterium
increase if that energy density is dark radiation, i.e., $\deltadr>0$.  This
can be remedied using a positive lepton number, as both \yp and \dtoh decrease,
while \neff and other neutrino cosmology parameters (i.e., the sum of the light
neutrino masses) are little changed by modest values of \lnustar (\neff and the
abundances scale differently for large values of \lnustar
\cite{2011JCAP...03..035M}).  However, the results of Secs.\ \ref{sec:dr} and
\ref{sec:w_deg} show that the two abundances do not have identical scaling
relations if \deltadr and \lnustar are to be used in concert.  In other words,
dark radiation and lepton asymmetry cannot together assuage tensions in
abundances.  Given the precision in measuring the primordial \dtoh observable,
changes of even a few percent could cause a strong tension.  The situation would
be even worse with precise measurements of $^3{\rm He}$ and $^7{\rm Li}$, as
the sign of the relative abundance changes is different compared to \yp and
\dtoh.  At this point, other physical mechanisms would need to be invoked to
mitigate internal BBN inconsistencies.  Returning to the \sinu scenario,
picking a mass of the $\varphi$ mediator for neutrino entropy creation during
BBN could change \yp and \dtoh such that the two scaling relationships are
identical when dark radiation is included.  Such a requirement on ${\rm SI}\nu$
--- or any other BSM scenario which must reconcile \yp and \dtoh with $N_{\rm
eff}$ --- will narrowly constrict the available parameter space.

%\clearpage 
\acknowledgments 

We thank Luke Johns, Amol Patwardhan, Andr\'e de Gouv\^ea, Yue Zhang, Walter Tangarife, and Julien Froustey for useful discussions. 
This research
was conducted through the Network for Neutrinos, Nuclear Astrophysics and
Symmetries collaboration and is supported by National Science Foundation, grant
PHY-1630782, and Heising-Simons Foundation, grant 2017-228. We also acknowledge
NSF grants PHY-1614864 and PHY-1914242 at University of California San Diego.  This research used resources provided by
the Los Alamos National Laboratory Institutional Computing Program, which is
supported by the U.S. Department of Energy National Nuclear Security
Administration under Contract No.\ 89233218CNA000001.

\bibliographystyle{JHEP}
\bibliography{master}

\end{document}